\documentclass[aps,prl,superscriptaddress,a4paper,floatfix,nofootinbib,twocolumn,10pt]{revtex4-1}
\usepackage{graphicx,color,amsmath,amssymb,enumerate}
\usepackage[normalem]{ulem}
\usepackage[ocgcolorlinks,colorlinks=true,linkcolor=blue,citecolor=red]{hyperref}

\newcommand{\tr}{\textrm{tr}}
\newcommand{\bra}[1]{\langle #1|}
\newcommand{\ket}[1]{|#1\rangle}

\begin{document}
\title{A short note on passivity, complete passivity and virtual temperatures}
\author{Paul Skrzypczyk}\affiliation{ICFO-Institut de Ciencies Fotoniques, Mediterranean Technology Park, 08860 Castelldefels, Barcelona, Spain}
\author{Ralph Silva}\affiliation{H. H. Wills Physics Laboratory, University of Bristol, Tyndall Avenue, Bristol, BS8 1TL, United Kingdom}
\author{Nicolas Brunner}\affiliation{D\'epartement de Physique Th\'eorique, Universit\'e de Gen\`eve, 1211 Gen\`eve, Switzerland}

\begin{abstract}
We give a simple and intuitive proof that the only states which are completely passive, i.e. those states from which work cannot be extracted even with infinitely many copies, are Gibbs states at positive temperatures. The proof makes use of the idea of virtual temperatures, i.e. the association of temperatures to pairs of energy levels (transitions). We show that (i) passive states are those where every transition is at a positive temperature, and (ii) completely passive states are those where every transition is at the same positive temperature.
\end{abstract}

\maketitle

\section{Introduction}
The notion of a passive state was introduced in the seminal work of Pusz and Woronowicz \cite{PusWor78} as a characterization of quantum states which cannot be processed to extract work. That is, given a state $\rho$ with Hamiltonian $H$, we ask whether the average energy can be lowered by a unitary transformation $U$ on the system, which is otherwise isolated (which is equivalent to a cyclic Hamiltonian process\footnote{that is a time-dependent interaction Hamiltonian $V(t)$ switched on only during $0 \leq t \leq \tau$.}). The change in average energy is denoted by $W$ and given by
\begin{equation}
W = \max_U \tr \left(H\left(\rho - U \rho U^\dagger\right)\right),
\end{equation}
States for which $W = 0$, i.e. states from which no work can be extracted, are referred to as passive states. On the other hand, states for which $W > 0$ are termed active states, and contain extractable work. 

It is possible to show that passivity can be re-expressed solely as a property of the state and its Hamiltonian. If we write $H = \sum_k E_k \ket{E_k}\bra{E_k}$  and $\rho = \sum_i \lambda_i \ket{i}\bra{i}$ in diagonal form, then $\rho$ is passive if and only if it satisfies the following two properties:
\begin{itemize}
\item $[\rho, H] = 0$, i.e. the state is (block) diagonal in the energy eigenbasis of the Hamiltonian. That is, the state eigenbasis $\{\ket{i}\}_i$ coincides with the energy eigenbasis $\{\ket{E_k}\}_k$. 
\item $E_i > E_j$ implies $\lambda_i < \lambda_j$, That is, the populations $\lambda_i$ of the energy levels are strictly decreasing as the energies $E_i$ increase.
\end{itemize}

In general one is not only interested in a single copy of a system, but in multiple copies that can be processed jointly. In particular, one may ask how much work can be extracted from $n$ copies of a state. Crucially, the composition of passive systems may not remain passive, hence exhibiting a form of activation. That is, there exist situations where a unitary $U$ acting on $n$ copies of a system is able to lower the average energy of the total system, whilst if one had access to only $n-1$ copies of the system no such unitary exists. This then naturally leads to the question of what is the class of states which remain passive under composition, i.e. from which no work can be extracted even from an infinite number of copies. Such states are termed completely passive. The celebrated result of \cite{PusWor78} is to show that the set of completely passive states is exactly equivalent to the set of thermal (or Gibbs) states
\begin{equation}
\rho = \frac{1}{Z}\exp(-\beta H),
\end{equation}
where $Z = \tr(\exp(-\beta H))$ is the partition function and $\beta = 1/k_BT$ is the inverse temperature. This result puts on firm grounds the notion of a heat bath in the form of an infinitely large Gibbs state, from which no work can be extracted. The importance of these results in the context of quantum thermodynamics has been highlighted, see for instance \cite{alicki,armin,fannes,karen,second}. 

This result was originally proven in the context of C$^*$ algebras and shortly afterwards translated into the framework of standard quantum mechanics by Lenard \cite{Len78}. In both cases, although the end result is intuitive, the proof does not convey too much intuition. In particular, it is not explicit why it is that passive but not thermal states can be activated for work extraction.

In this short note we provide a simple and intuitive proof based upon the idea of associating temperatures to pairs of energy levels. This is the idea behind the concept of virtual temperatures and virtual qubits introduced in \cite{BruLinPop12}, and further discussed in \cite{luis,paul}. Furthermore, our technique illustrates explicitly how work can be extracted from a sufficient number of copies of a passive but non-thermal state, hence giving an upper bound on the number of copies needed. 

\section{Virtual temperatures}
We start by discussing in more detail the idea of assigning a temperature to a pair of levels which from here on out we will find convenient to refer to as a \emph{transition}. Consider a system $\rho$ with Hamiltonian $H = \sum_k E_k \ket{k}\bra{k} $ comprised of $d$ energy eigenstates $\ket{k}$, with energy eigenvalues $E_k$, and assume $\rho$ is diagonal in this basis, $\rho = \sum_k \lambda_k \ket{k}\bra{k}$. In total there are  $d(d-1)/2$ transitions between energy levels. Consider the transition between the energy states $\ket{i}$ and $\ket{j}$, and let us assume without loss of generality that the gap $E_i - E_j > 0$\footnote{We do not define virtual temperatures for degenerate transitions. Since no work can ever be extracted from such transitions we will never need such a concept.}. Given populations $\lambda_i$ and $\lambda_j$ respectively, we associate a virtual inverse temperature $\beta_v$ to the transition, given by
\begin{equation}\label{e:betav def}
	\beta_v = \frac{\log \lambda_j - \log \lambda_i}{E_i - E_j},
\end{equation}
which arises by identifying the ratio of the populations with the Boltzmann factor, $\lambda_i/\lambda_j = e^{-\beta_v(E_i-E_j)}$. First we note that in general the virtual inverse temperature defined this way need not be positive. In particular, whenever $\lambda_i > \lambda_j$, i.e. the transition has a population inversion, then the temperature will be negative. Second, the reason for calling this a virtual temperature is because by coupling an external system to this transition, one can prepare physical systems at the virtual temperature, i.e. it behaves in this respect like a real temperature \cite{BruLinPop12}. 

It will also be important to us to understand how virtual temperatures transform under composition.  To that end, consider now two systems. For the first system we consider the transition between states $\ket{i}_1$ and $\ket{j}_1$, with energy gap $\Delta_1 = E_i - E_j$, populations $\lambda_i$ and $\lambda_j$ and virtual inverse temperature $\beta_v^1$. Likewise for the second system we consider transitions between the states $\ket{i'}_2$ and $\ket{j'}_2$, with gap $\Delta_2 = E_{i'} - E_{j'}$, populations $\lambda_{i'}$, $\lambda_{j'}$ and virtual temperature $\beta_v^2$. 

Now, the joint system features 2 non-trivial transitions (i) between the pair of levels $\ket{i}_1\ket{j'}_2$ and $\ket{j}_1\ket{i'}_2$, and (ii) between $\ket{i}_1\ket{i'}_2$ and $\ket{j}_1\ket{j'}_2$. For case (i), the population of the first level is $\lambda_i\lambda_{j'}$, the second $\lambda_j\lambda_{i'}$, and the gap  is $\tilde{\Delta} = \Delta_2 - \Delta_1$ (where we have assumed without loss of generality that $\Delta_2 > \Delta_1$). From \eqref{e:betav def} the inverse virtual temperature $\tilde{\beta}_v$ of the composed transition is given by
\begin{align}\label{e:comp}
\tilde{\beta}_v &=  \frac{\log (\lambda_i\lambda_{j'}) - \log (\lambda_j\lambda_{i'})}{\Delta_2 - \Delta_1} \nonumber \\
&=  \frac{(\log\lambda_{j'} - \log\lambda_{i'}) - (\log \lambda_j-\log\lambda_{i})}{\Delta_2 - \Delta_1} \nonumber \\
&= \frac{\beta_v^2\Delta_2 - \beta_v^1\Delta_1}{\Delta_2 - \Delta_1}
\end{align}
where we have used equation \eqref{e:betav def} in the final step. Note that we obtain the same expression in the case $\Delta_1 > \Delta_2$ (i.e. the formula is insensitive to the sign of the gap, and so in fact we did not need the assumption $\Delta_2 > \Delta_1$). For case (ii) the populations are now $\lambda_i\lambda_{i'}$ and $\lambda_j\lambda_{j'}$, the gap is $\tilde{\Delta}' = \Delta_2 + \Delta_1$, and a similar analysis shows that the inverse virtual temperature $\tilde{\beta}'_v$ of the transition is given by
\begin{align}\label{e:comp2}
\tilde{\beta}'_v &=  \frac{\beta_v^2\Delta_2 + \beta_v^1\Delta_1}{\Delta_2 + \Delta_1}
\end{align}
Hence we see that the inverse virtual temperatures compose linearly in both cases. Finally, we note that $\tilde{\beta}'_v$ is always in between the composed temperatures $\beta_v^1$ and $\beta_v^2$, while $\tilde{\beta}_v$ is in fact always larger than the biggest or smaller than the smallest temperature. 

\section{Passivity}
The notions introduced above will now allow us to re-express the notion of passivity for diagonal states in simple terms\footnote{It would be interesting to extend the notion of virtual temperatures to non-diagonal states. This is however work in progress, thus beyond the scope of this short note}. Specifically the second requirement for a state to be passive, i.e. that $E_i > E_j$ implies $\lambda_i < \lambda_j$, is equivalent in the language of virtual temperatures to the requirement that the virtual temperature of every transition is positive. One the one hand if the state is ordered in population then by construction all of the virtual temperatures are positive. On the contrary, if the state has one or more negative virtual temperatures, work can be extracted by exploiting the associated population inversion. In particular, assume that the levels $\ket{i}$ and $\ket{j}$ with $E_i - E_j > 0$ have a virtual temperature $\beta_v < 0$. Consider then the following unitary transformation on the system
\begin{equation}\label{e:U}
U = \ket{i}\bra{j} + \ket{j}\bra{i} - \ket{i}\bra{i} - \ket{j}\bra{j} + \openone.
\end{equation}
The amount of work extracted from the system (the change in its average energy) is given by
\begin{align}
W &= (E_i - E_j)(\lambda_i - \lambda_j) \nonumber \\
&= (E_i - E_j)\lambda_j(1-e^{-\beta_v(E_i-E_j)})
\end{align}
which must be non-positive for the state to be passive. However, $(E_i - E_j)\lambda_j$ is positive by construction, and $(1-e^{-\beta_v(E_i-E_j)}) > 0$ whenever $\beta_v < 0$. Thus $W > 0$, and the state is non-passive if it contains a negative virtual temperature.  

For completeness, in the appendix we show that all virtual temperatures being positive implies that no work can be extracted from the state.  We defer this to the appendix since it is essentially a reproduction of Lenard's proof \cite{Len78} that does not rely on virtual temperatures, and is not necessary for the discussion in the next section regarding complete passivity.
\section{Complete passivity}
We have seen above that passive states are those where every transition is at a positive temperature. We are now going to show that completely passive states are those where every transition is at the same positive temperature\footnote{There is one exceptional case, when the system has a degenerate ground-space; then any ground-state is also completely passive (but need not be thermal). }. The proof works by showing that whenever a system has two (or more) transitions at different virtual temperatures, then by composing sufficiently many copies, the combined system always has a transition at a negative temperature, and is therefore not passive. 

Consider again a system $\rho$ with $d$ levels and consider first a transition between states $\ket{i}$ and $\ket{j}$, with gap $\Delta_1 = E_i - E_j$ and virtual inverse temperature $\beta_v^1$. Let us consider $n$ copies of $\rho$, and the same transition in each system. Now, by applying the composition rule \eqref{e:comp2} $n-1$ times, it is straightforward to see that the joint system has a transition between the states $\ket{i}^{\otimes n}$ and $\ket{j}^{\otimes n}$ with gap $n \Delta_1$ and the same virtual inverse temperature $\beta_v^1$. Similarly, consider another transition of $\rho$ between the states $\ket{i^\prime}$ and $\ket{j^\prime}$, with gap $\Delta_2 = E_{i'} - E_{j'}$ and virtual inverse temperature $\beta_v^2 > \beta_v^1$, without loss of generality. Now consider another $k$ copies of $\rho$. Exactly as above, the joint system  has a transition between the states $\ket{i^\prime}^{\otimes k}$ and $\ket{j^\prime}^{\otimes k}$ with gap $k\Delta_2$ and the virtual inverse temperature $\beta_v^2$.

Finally, for the $n+k$ copies of $\rho$ together, consider the transition between the states $\ket{i}^{\otimes n}\ket{j^\prime}^{\otimes k}$ and $\ket{j}^{\otimes n} \ket{i^\prime}^{\otimes k}$. This transition has an energy gap of $n\Delta_1 - k\Delta_2$, and from the composition rule \eqref{e:comp}, it has the inverse virtual temperature
\begin{align}\label{e:bv}
\beta_v &= \frac{\beta_v^1 n\Delta_1 - \beta_v^2 k\Delta_2}{n\Delta_1 - k\Delta_2} 
\end{align}
Now, since $\beta_v^2 > \beta_v^1$, i.e. the two transitions are at different virtual temperatures, then it is always possible to find a negative $\beta_v$, by choosing an appropriate number of copies $n$ and $k$ such that the numerator of \eqref{e:bv} is negative, whilst the denominator is positive. In particular, choosing 
\begin{equation}\label{e:crit}
\frac{\Delta_2}{\Delta_1} < \frac{n}{k} < \frac{\Delta_2}{\Delta_1}\frac{\beta_v^2}{\beta_v^1}
\end{equation}
ensures that $\beta_v<0$. Therefore, by applying a unitary of the form \eqref{e:U} but now with $\ket{i}$ replaced by $\ket{i}^{\otimes n} \ket{j^\prime}^{\otimes k}$ and $\ket{j}$  replaced by $\ket{j}^{\otimes n} \ket{i^\prime}^{\otimes k}$ work will be extracted from the combined system of $n+k$ copies. Thus, any passive state containing two or more virtual temperatures is not completely passive. The only possibility for a completely passive state is thus one containing only a single virtual temperature, which is precisely the defining property of a thermal state.

Finally, it is worth noting that \eqref{e:crit} provides a sufficient condition for finitely many copies of a state to become non-passive: One has to find the smallest number $n+k$ such that \eqref{e:crit} is satisfied for any possible pair of virtual temperatures in the system. Note that a similar bound was also obtained considering almost deterministic work extraction in \cite{second}. 

\section{Conclusions}
In summary, we have presented what we believe is a simple and insightful alternative proof that the only completely passive states are thermal states. The only notion that our proof relies upon is the association of virtual temperatures to transitions in a system via Gibbs weights, and can be simply stated as the only completely passive states are those which contain a single virtual temperature. To show this we proved that every passive but not completely passive state has the property that upon composition one can find a transition at a negative virtual temperature, from which work can then be extracted. This statement is intuitive from the perspective of thermal machines, where having access to two baths at differing temperatures is all that is required to build a work extracting machine. 

\emph{Acknowledgements.} We thank Markus M\"uller and Marcus Huber discussions. We acknowledge financial support from the Swiss National Science Foundation (grant PP00P2\_138917) and an EU Marie Curie COFUND action (ICFOnest).

\begin{appendix}
\section{Appendix}
In this appendix, reproducing essentially the proof of Lenard \cite{Len78} we show that whenever a state is such that all of the virtual temperatures are positive then no work can be extracted from the state, i.e. this shows that all virtual temperatures being positive is also a sufficient condition for the inability to extract work. 

To be explicit, let us consider a diagonal state $\rho$ such that $\rho = \sum_k \lambda_k \ket{E_k}\bra{E_k}$, where $\ket{E_k}$ are energy eigenstates (of the Hamiltonian $H = \sum_k E_k \ket{E_k}\bra{E_k}$) such that $E_{k+1} \geq E_k$, and $\lambda_{k+1} \leq \lambda_k$. This state is the most general diagonal state which has all of its virtual temperatures positive. 

The average energy of this state is given by
\begin{equation}
\langle H \rangle = \tr\left(H\rho\right) = \sum_k \lambda_k E_k.
\end{equation}
Let us now consider that we apply an arbitrary unitary $U = \sum_{ij} u_{ij}\ket{E_i}\bra{E_j}$ with which we would like to reduce the average energy. We see that the average energy of the final state is
\begin{equation}\label{e:final energy}
\langle H\rangle ' = \tr\left(H U\rho U^\dagger\right) = \sum_{ij} \lambda_j E_i |u_{ij}|^2
\end{equation}
Now, if we define the matrix $S = \sum_{ij}|u_{ij}|^2 \ket{E_i}\bra{E_j}$, then due to the unitarity of $U$, it follows immediately that $S$ is a doubly-stochastic matrix, i.e. that the sum of each row and of each column of $S$ is unity. Moreover, by varying over all unitary matrices we can generate all doubly-stochastic matrices in this way. Thus in order to find the unitary $U$ that minimises the average energy of $U\rho U^\dagger$ it suffices to find the doubly-stochastic matrix which minimises the expression \eqref{e:final energy}.

Finally, we use the fact that \eqref{e:final energy} is in fact a linear function of the stochastic matrix $S$. This is important, since the minimum of a linear function over a convex set (here the convex set of doubly-stochastic matrices) always occurs at one of the extreme points of the set. In the present context, the convex set of doubly-stochastic matrices is in fact a polytope (a convex set with finitely many extreme points, or vertices), with the vertices exactly the permutation matrices.

What this shows is that the unitary which minimises the average energy leaves the state $\rho$ diagonal in the energy eigenbasis, and only interchanges populations between energy levels, i.e. one can never decrease the average energy by creating coherences between energy levels. Finally, it is straightforward to see that all permutations on the state ordered in energy either leave the average energy unchanged (if the permutation is only between levels with the same energy), or increase the energy. Thus this state is passive, as no work can be extracted from it. 

We end finally by noting that the above analysis also serves to show that any non-diagonal initial state necessarily is non-passive, unless the coherence appears in a degenerate subspace. In particular, consider that $\sigma$ is the non-diagonal state (with coherences between states that differ in energy), and $\rho = V\sigma V^\dagger$ is the state diagonal in the energy eigenbasis ordered in energy. The doubly-stochastic matrix associated to $V$ is definitely not a permutation matrix, since $\sigma$ and $\rho$ are not diagonal in the same basis. Thus we see that $\sigma$ will have larger energy than $\rho$, since only permutation matrices minimise the expression \eqref{e:final energy}. 
\end{appendix}
\end{document}